# Superconductivity in boron-doped carbon nanotube networks

*Jie Pan*[1†], *Bing Zhang*[1†], *Yuxiao Hou*[1†], *Ting Zhang*[1], *Xiaohui Deng*[1], *Yibo Wang*[1], *Ning Wang*[1], *Ping Sheng*[1,2*]

[1]*Department of Physics, Hong Kong University of Science and Technology, Clear Water Bay, Kowloon, Hong Kong, China*

[2]*Institute for Advanced Study, Hong Kong University of Science and Technology, Clear Water Bay, Kowloon, Hong Kong, China*

[†] These authors contribute equally to this work.

[*]Correspondence author. Email: sheng@ust.hk

## Abstract

By using the 5-Angstrom diameter pores of calcined zeolite ZSM-5 as the template, we have fabricated boron-doped carbon nanotube networks via the chemical vapour deposition method. Raman data indicate the network to comprise segments of interconnected (3,0) carbon nanotubes. Four-probe transport measurements showed a superconducting transition initiating at 40 K, with a sharp downturn at ~20 K to a low resistance state at 2 K, accompanied by a low resistance plateau in the *I-V* characteristic, fluctuating around zero resistance. The magnetic SQUID measurements exhibited the Meissner effect characteristic of thin superconducting wire networks in which the wire radius $a$ is much smaller than the London penetration length $\lambda_L$. At low magnetic field, the negative diamagnetic susceptibility was observed to persist beyond 200 K. The transport and magnetic data are reconciled on the basis of a physical model based on weak links comprising short, one-dimensional (1D) superconducting nanotubes, that govern the global transport behaviour. Fitting of the transport data by theory, based on the mechanism of phase slips in 1D superconductors, yields an intrinsic superconducting $T_C$~200 K, consistent with the magnetic susceptibility data. This high intrinsic transition temperature is further reinforced by the appearance of the weak link's 1D metastable states at the intrinsic $T_C$, attendant with their contribution to the transport noise that arises from the thermally activated phase slips.

# I. Introduction

There has been a continuing effort on carbon nanotube superconductivity since its early observations two decades ago [1-13], sustained by the prospect establishing a superconducting carbon system with 1D elements. Recently, by using calcined zeolite ZSM-5 as a template, networks of carbon nanotubes were grown in the 5-Angstrom diameter pores via the chemical vapour deposition (CVD) of methane at 800 °C. Four probe measurements of both the transverse and longitudinal resistances indicated a Peierls type metal-insulator transition at 30 K [14]. Peierls transition is driven by the electron-phonon (ep) interaction; its observation in the nanotube@ZSM-5 samples is aligned with the theoretical prediction of increasing ep coupling with decreasing nanotube diameter [15]. Indeed, Raman spectroscopy of the samples yielded a radial breathing mode (RBM) frequency in the close vicinity of (3,0) carbon nanotubes, with a diameter of ~0.3 nm.

Carbon-based nanostructures [16,17] have been regarded as good candidates for high $T_C$ superconductivity due to the high Debye temperature and light atomic weight. Some carbon materials become superconducting by chemical doping, such as potassium-doped $C_{60}$ [18] and alkali-metal-intercalated graphite [19-21]. In contrast to the easily oxidized alkali metals, boron doping represents a much more stable method. Diamond [22] and (10,0) carbon nanotubes (diameter ~ 1 nm) [23] were found to be superconducting after boron doping. In this work, we report evidences of superconductivity in boron-doped (3,0) carbon nanotubes networks, evidenced by both electric transport and magnetic SQUID measurements. The latter points to a transition temperature in excess of 200 K.

It is well known that strong ep coupling is an important ingredient for superconductivity. Its presence in nanotube@ZSM-5 [14] implies that it is a favourable system to induce superconductivity through boron doping. We used diborane gas to boron-dope the (3,0) carbon nanotube networks during the CVD process via methane decomposition. Transport and magnetic SQUID measurements of the resulting samples yielded evidences of superconductivity, albeit with important differences with the usual bulk superconductors. In particular, the Meissner effect displayed the characteristics pertaining to those of thin superconducting wire networks with wire diameters much smaller than the magnetic penetration length.

## II. Statement of main results

Below we first give the physical characteristics of the zeolite template and its relevant pore geometry in Fig. 1, followed by a description of the CVD process and the method of generating the diborane gas for doping. Raman characterization results are presented in Fig. 2, followed by a description of the fabrication process for the device used in the transport measurements. It is shown that the four-probe measurements showed a superconducting transition initiating at ~40 K, with a sharp downturn at ~20 K to a low resistance state at 2 K. This is shown in Fig. 3(a). *I-V* characteristics displayed a low resistance gap in Fig. 3(b), accompanied by fluctuations around the low resistance plateau. The low resistance gap has a small 40 nA critical current, that became smeared with increasing temperature and disappeared above 40 K , as shown in Fig. 3(c). Fluctuations and noise in transport measurements are summarized in Fig. 4. We substantiate the evidence for superconductivity by carrying out the magnetic SQUID measurements. Clear diamagnetic susceptibility was observed, shown in Fig. 5(a). Its temperature dependence, and the magnitude of the susceptibility, are in excellent agreement with the theoretical and simulation prediction for a thin superconducting wire network with wire diameter much smaller than the magnetic penetration length $\lambda_L$ as shown in in Fig. 5(b). At low applied magnetic field, the diamagnetic susceptibility was seen to persist up to a high temperature, in excess of 200 K. Below we show that the two seemingly incompatible sets of data, one on transport and one on the magnetic susceptibility, can be reconciled on the basis of the following physical picture.

While the uniform three-dimensional (3D) networks of carbon nanotubes would be an ideal structure as shown in Fig. 1(a), yet due to the inhomogeneous nature of our sample, a plausible scenario is that there can exist regions where a superconducting nanotube is not connected to any side branches of other nanotubes, i.e., within a stretch that can persist over several unit cells long, the relevant superconducting nanotube represents a true one-dimensional (1D) superconductor. These regions serve as the weak links that dominate the global transport superconducting behaviour. A 1D superconductor is characterized by metastable current-carrying states with energy barriers separating the neighboring states [24]. The energy barrier height is directly proportional to the cross-sectional area of the 1D superconductor. Transition between two metastable states is denoted a "phase slip" that can generate a voltage and contribute to transport noise. Since in our case this energy barrier against phase slips can be very small, hence it is expected that the observed transition temperature to a resistive state of the 1D weak links should be relatively low. This low phase slip energy barrier ensures that globally, transport superconducting behaviour can only be observed at a much lower temperature than its intrinsic $T_C$. Moreover, from the theory of Langer-Ambegoakar-McCumber-Halperin (LAMH) [25,26], the critical current of the weak links is controlled by the condition $2\pi n/\ell < 1/\sqrt{3}$ for the observation of superconductivity [9,24], where the right-hand side of the inequality denotes the limit beyond which the 1D superconductor turns normal. Here $\ell$ is the length of the 1D weak link (in unit of the coherence length), and $n$ is an integer (the winding number), denoting the ordering of the metastable current-carrying states, with a larger $n$ implying a larger current. If $\ell$ is not too large compared to the coherence length, then $n$ needs to be small, implying a small critical current. This was indeed observed in our sample.

It follows from the above that the transport data is dominated by the behaviour of the 1D superconducting weak links, whereas the magnetic data represent the behaviour for the rest 3D, inter-connected part of the network. Below we show that the above physical picture can be quantitatively substantiated by the theory prediction of resistance versus temperature resulting from phase slips [13,24,27-30]. Fitting to the transport data yields an intrinsic $T_C \sim 200$ K, i.e., consistent with that obtained from the magnetic data. This high intrinsic $T_C$ is further reinforced by the jump of the transport noise variance at a temperature >200 K, owing to the appearance of the metastable states in the weak links below the intrinsic transition temperature, with the attendant thermally excited phase slips.

As the present approach represents a relatively reliable method of obtaining superconductivity in carbon nanotube networks, it holds the potential for further developments.

### III. Sample fabrication and Raman characterization

The calcined zeolite ZSM-5 was used as a template for growing carbon nanotube networks. The ZSM-5 has an Angstrom-scale porous nanostructure. In Fig. 1(a) we show the Scanning Electron Microscope (SEM) image of a batch of ZSM-5 zeolite crystallites. The typical size of a ZSM-5 zeolite crystallite ranges from 3 to 4 microns, sufficiently large for nano-fabrication and electronic transport measurements. In Fig. 1(b) we show a schematic picture of the pore structure. Along the *a* axis, there are straight channels with an inner diameter around 5 Angstroms. Along the *b* axis, there exist undulating channels with a similar inner diameter. All the channels in ZSM-5 are inter-connected, with ~1 nm segments, to form a three-dimensional (3D) network. The carbon nanostructures grew inside the ZSM-5 template are thus expected to be a 3D network comprising 1D carbon nanotubes.

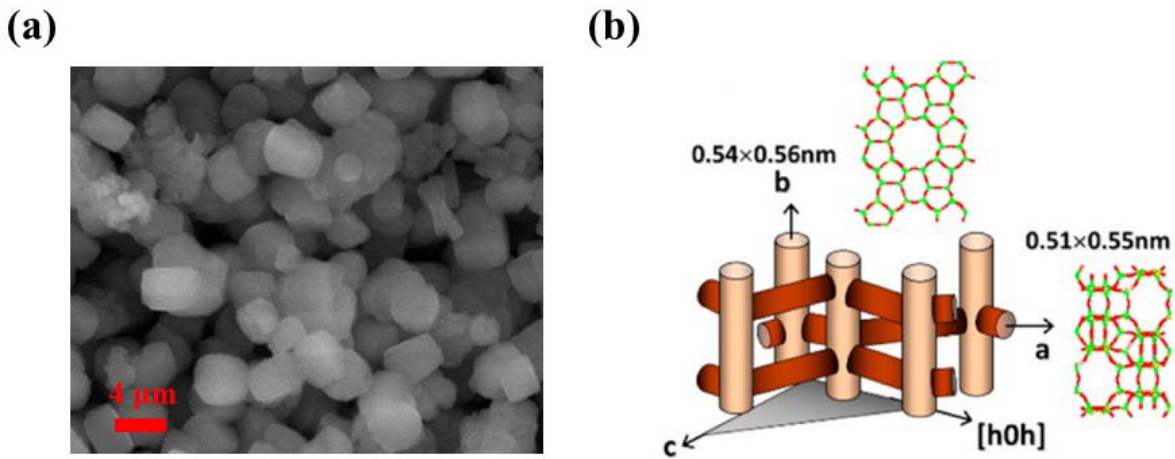

**Figure 1 (a)** SEM image of the ZSM-5 zeolite crystallites, each ranges from 3 to 4 microns across. They serve as the template for growing boron-doped networks of carbon nanotubes inside their pores. **(b)** A schematic picture of the pore structure of ZSM-5 along the three axes.

Boron-doped carbon nanotube networks were grown in ZSM-5 by using the CVD method in which the methane and diborane gas were mixed together and heated under an elevated pressure and temperature. We generated the diborane gas, $B_2H_6$, used in the doping by reacting iodine with sodium borohydride *in-situ*, via the reaction $I_2 + 2NaBH_4 \rightarrow 2NaI + B_2H_6 + H_2$, inside a vessel that was initially pumped to vacuum. Then the vessel was heated to 180°C and the diborane gas was slowly generated to reach 1.2 atmospheres after ~2 hours. Before the growth of boron-doped carbon nanotubes, the calcined ZSM-5 was heated to 200°C in vacuum for 2 hours to remove the water vapour adsorbed in the ZSM-5 pores. This step happens to be very important since otherwise the diborane gas would be dissipated by reacting strongly with the released water vapour. After drying, the diborane gas (~8 cm³ at 1.2 atmospheres of $B_2H_6 + H_2$) ) and methane (~28 cm³ at ~ 6 atmospheres) were released into the CVD chamber in the presence of the calcined ZSM-5; the total pressure was kept at 6 atmospheres, with a temperature maintained at 800 °C for 10 hours.

To characterize the sample, Raman measurement was conducted with an excitation laser wavelength of 514 nm. The measured Raman spectrum is shown in Fig. 2. The prominent G peak (~1600 cm$^{-1}$) and D peak (~1390 cm$^{-1}$) are typical for sp2 carbon bonds [31,32]. There exists an RBM peak at 805 cm$^{-1}$, which is very close to the theoretically predicted RBM of (3,0) carbon nanotubes at 817 cm$^{-1}$. The diameter of the carbon nanotubes is estimated to be 0.29 nm [14]. The closest alternative, the RBM for (2,1), is calculated to be 10% higher. The peak around 1190 cm$^{-1}$ is known as the intermediate frequency modes (IFM), which represent the combination of both RBM and D peaks [33,34].

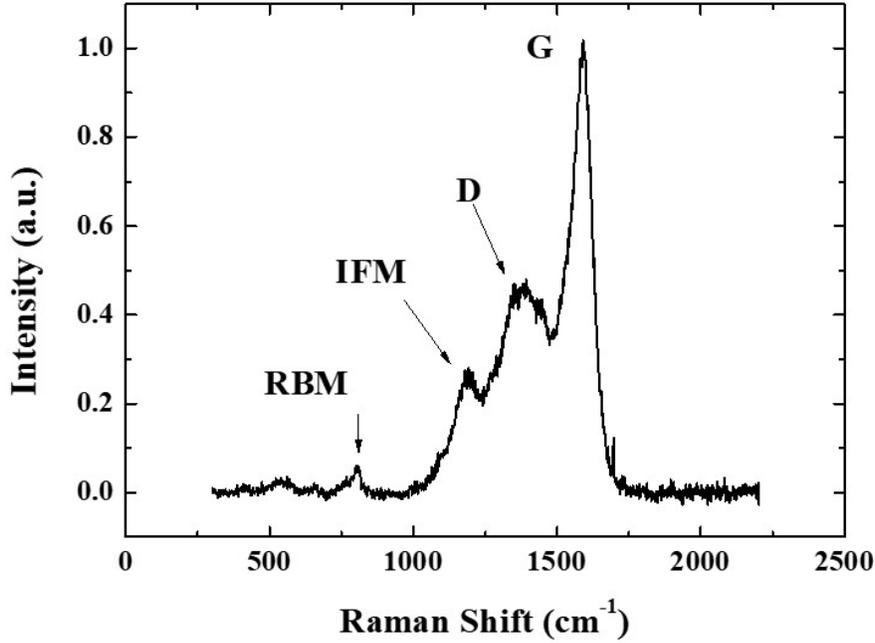

**Figure 2** Raman spectrum of boron-doped sample of carbon nanotube networks embedded in ZSM-5 zeolites.

To investigate the electronic transport properties of boron-doped carbon nanotube networks, devices were fabricated by first transferring the ZSM-5 zeolite crystallites onto a quartz substrate covered with photoresist (PMMA A9) in order to fix the small crystallites. A thin 5 nm adhesive layer of Ti was first sputtered on the whole spread, followed by 60 nm of Au. Subsequently the focus ion beam (FIB) was used to etch a chosen crystallite to form the electrodes. At the edge of the etched crystallite, Pt was deposited by FIB to ensure good contact to the device. The detailed nanofabrication technique is detailed in Ref [6]. An SEM image of the four-probe device is shown in the inset of Fig. 3(a). The separation between the two voltage probes is 300 nm.

## IV. Results
### (a) *Electrical Transport*

The device was transferred into the Physical Properties Measurement System (PPMS Quantum Design) for electronic transport measurements. We use the DC method to measure the resistance. The voltage nanometer Keithley 2182A was used for detecting the four-probe voltage drops. Two sources were used; one is the current source Keithley 6221 for directly applying the DC bias current and the other is the voltage source Keithley 2612B for applying the DC bias voltage. The bias current and bias voltage were measured simultaneously. These two measurement methods were both found to be reliable. At a fixed temperature, the *I-V* characteristic is scanned and the resistance is fitted by selecting data within 60 nA unless specified otherwise.

The temperature dependence of resistance for both the boron-doped and un-doped case (the latter was fabricated in exactly the same manner, but without the diborane gas) is shown in Fig. 3(a). For the un-doped case (the black curve), the resistance exhibited an upturn as temperature decreased below ~ 30 K owing to the Peierls-type transition at low temperatures, which has been detailed in Ref [14]. In contrast, for the boron-doped case, the 4-probe resistance $R$ (red dots) started to drop below ~ 40 K, followed by a sharp downward turn at ~20 K to a low resistance state at ~ 2 K. The contrasting temperature dependencies between the two samples illustrate the effect of boron doping in altering the electronic properties of the networks of carbon nanotubes. With the enhancement of carrier density after boron-doping, the superconducting behaviour prevails.

It is noted that the resistance data of the doped sample are higher than that of the un-doped sample in almost all the temperature range, up to 200 K. This might seem puzzling, since intuition would put it the other way around. Here we adopt a heuristic explanation, i.e. it is attributed to the scattering and localization effect in 1D systems. In the normal state, the 1D weak link is just a 1D metallic wire, with doped boron atoms acting as added scattering centres. Since in 1D the Anderson localization length is proportional to the scattering mean free path, a decreased mean free path can imply an increased sample resistance in an exponential manner.

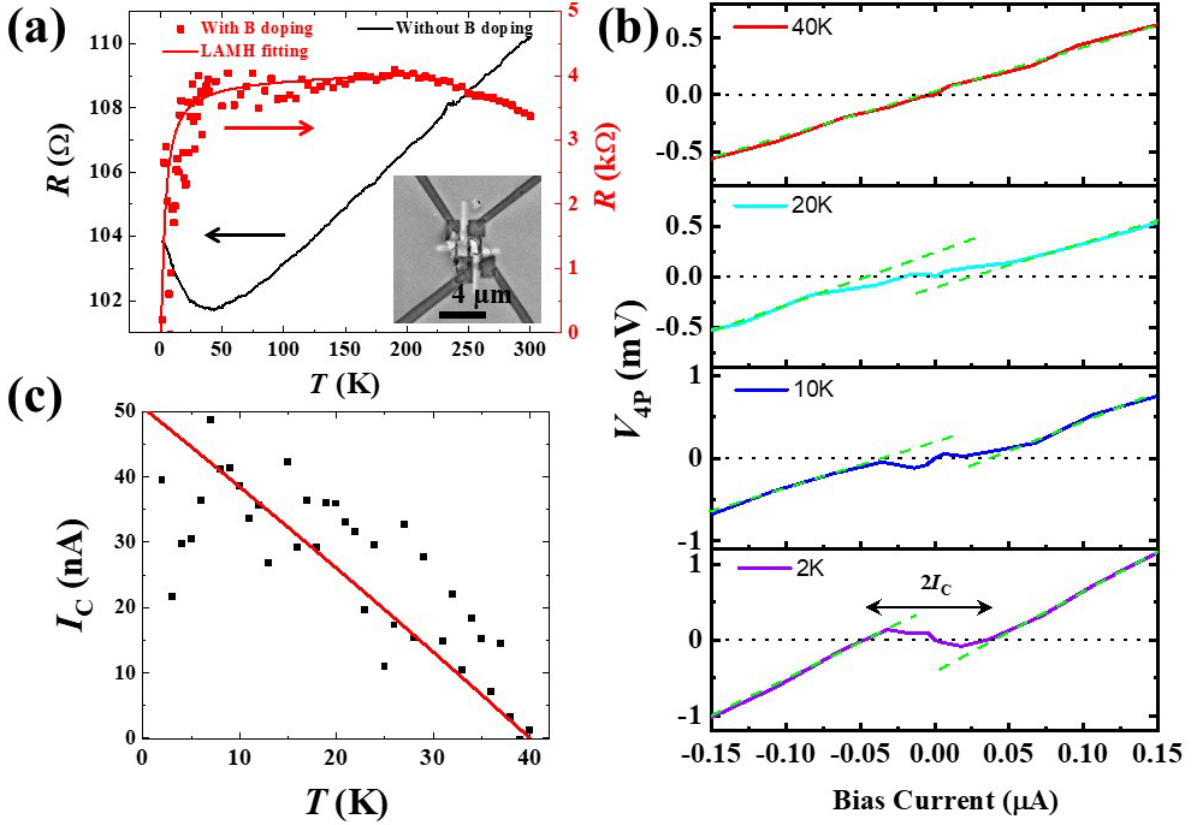

**Figure 3 (a)** Temperature dependence of resistance for boron-doped case (red squares) fitted by the LAMH theory (red solid line) and without boron-doping case (black solid line). The inset is an SEM image of the fabricated device. **(b)** *I-V* characteristics measured at 2, 10, 20 and 40 K. The green dashed curves denote the linear *I-V* curves beyond the critical currents, and the thin black curves exhibit the zero bias. **(c)** Critical current $I_C$ summarized as a function of temperature, shown by the black squares. Red curve denotes the fitted result.

The temperature dependence of resistance can be fitted by using following formula [13] derived from the LAMH theory [25,26] for the 1D superconductors:

$$R = R_N \exp\left[-A\frac{T_C}{T}\left(1+\frac{T}{T_C}\right)^2\left(1-\frac{T}{T_C}\right)^{3/2}\right] \quad , \tag{1}$$

where $R_N$ is the normal state resistance at the intrinsic critical temperature $T_C$, and $A$ is a dimensionless parameter. The product $AT_C$ is indicative of the phase slip energy barrier, which is proportional to the cross sectional area of the carbon nanotubes. From the results on 4-Angstrom carbon nanotubes, $AT_C$ was found to be around 5.4 K [13]. It follows that for our 3-Angstrom (3,0) carbon nanotube, $AT_C$ should be ~ 3 K. Based on this constraint, the best fitting to the temperature variation of resistance shown in Fig. 3(a) yields $A = 0.015$ and $T_C = 200$ K, as delineated by the solid red curve. It should be noted that owing to the small phase slip energy barrier in the ultrathin carbon nanotube, the sharp downturn of resistance appeared at ~ 20 K, much lower than the $T_C$ = 200 K obtained from fitting to the data. Below we show that such a high intrinsic critical temperature is consistent with the data on the variance of the transport noise, as well as with the magnetic SQUID measurement results.

The measured *I-V* curves are presented in Fig. 3(b). At low temperatures, the *I-V* curves are featured by a low resistance plateau around the zero bias current. It is seen that there exists a critical current $I_C$ beyond which the low resistance plateau disappears. As temperature increases, the critical current decreases and the nonlinear behaviour first become smeared and then disappeared above 40 K. In Fig. 3(c), we summarize the critical current measured at different temperatures from 2 K to 40 K, shown by the black squares. This temperature dependence of the critical current can be fitted by using the formula, $I_C \sim I_0\left(\sqrt{1-(T/T_C)} - 2/\sqrt{5}\right)$ as derived in Appendix C, where $T_C$ = 200 K is given by fitting *R-T* relations with Eq. (1), shown by the red curve. The small zero temperature critical current, ~ 50 nA, is interpreted to result from the relatively short 1D superconducting weak link (in unit of the coherence length) within the framework of the physical picture delineated above. The fact that the critical current vanishes at 40 K, rather than at $T_C$ = 200 K, is due to the discreteness of *n* for the cut-off critical current.

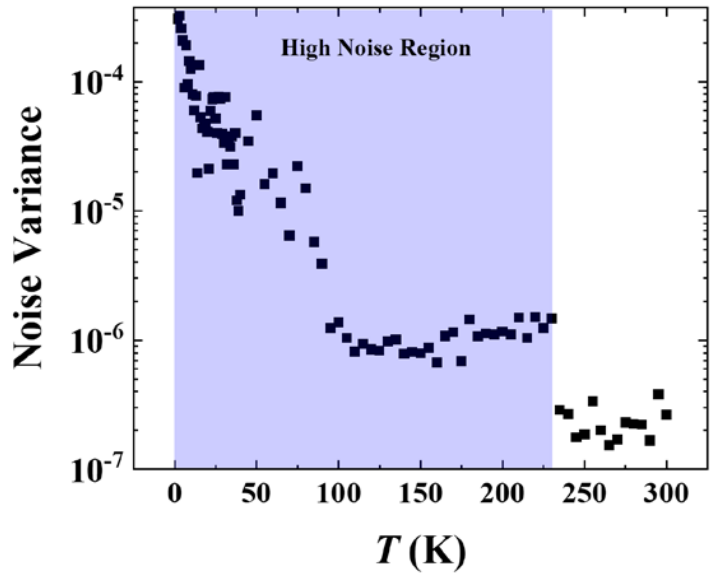

**Figure 4** Measured transport noise variance plotted as a function of temperature. It is seen that there is a clear jump in the variance value at around 230 K, signaling the appearance of the metastable states and the attendant phase slips

below this intrinsic transition temperature. Such behavior has been observed in other 1D superconductivity sample (unpublished), but at a much lower intrinsic transition temperature.

The resistance data of the doped sample are seen to display large scatter at low temperatures, this fluctuation can be clearly seen in Fig. 3. We attribute such scatter to the random, time-varying fluctuations of the resistance caused by thermally excited phase slips that can cause sudden changes in the current/voltage. The thermal excited phase slips above 40 K would be mostly from the metastable states to the normal state of the weak link. Such thermally-excited resistance noise is evidenced by a sudden jump in the measured variance of the resistance noise below 230 K, as shown in Fig. 4. The noise is calculated by the definition as

$$\text{Noise Variance}(T) = \frac{\sum_{|I_j|<60nA}\left(V_{\exp}(I_j,T)-V_{\text{fit}}(I_j,T)\right)^2}{V_{\text{fit}}(60\text{ nA},T)^2}, \qquad (2)$$

where $V_{\exp}$ is measured voltage drop and $V_{\text{fit}}$ is the linearly fitted voltage drop. We conclude that the observed noise is caused by the 1D weak links. On the other hand, the increase in the noise variance at low temperatures is mostly likely due to the growth of the coherence length as the low (zero) resistance state is approached.

### (b) *Magnetic susceptibility*

The Meissner effect for a network of very thin superconducting wires must be different from that of the bulk sample, owing to the fact that the thin superconducting wires cannot completely expel the magnetic field. In order to obtain an a-priori physical picture of what should be the expected behavior, we have carried out both analytical solution of the equation,

$$\nabla^2 H - \frac{1}{(\lambda_L)^2} H = 0, \qquad (3a)$$

for a single cylindrical wire with radius *a*, for the two cases where **H** is either parallel or perpendicular to the axial direction of the wire, as well as simulations for a simple cubic lattice of such wires with a lattice constant of 4*a*. We have also carried out simulations on a network of superconducting wires, to model the effect of a three-dimensional simple cubic lattice of (3,0) carbon nanotubes by a simple cubic lattice of intersecting superconducting cylinders with a lattice constant of 1 nm and a cylinder diameter of 0.5 nm. A schematic illustration of the model is shown as inset in Fig. 5(b). For the purpose of simulation, the lateral *xy* directions are periodically repeated, and along the *z* direction the model has 10 unit cells, ranging from *z*=0 to 10 nm. A 1 Tesla magnetic field is applied from the negative *z* direction. The equation to be solved in the COMSOL is given by the London equation

$$\nabla^2 \mathbf{H} - \frac{1}{\lambda_L^2}\mathbf{H} = 0 \text{ , with } \lambda_L = \begin{cases} 1 \sim 10 \text{ nm} & : \text{inside the cylinder} \\ 10^8 \text{ nm} & : \text{elsewhere} \end{cases}. \tag{3b}$$

The boundary condition at the interface is assumed to be continuity for both the parallel and perpendicular cases. Details of the analytical solution are given in the Appendix B. The results show that for a single wire, the analytic solution of the magnetic susceptibility $\chi$ is given by

$$\chi_{\text{single}} = \frac{2\lambda_L}{a}\frac{I_1(a/\lambda_L)}{I_0(a/\lambda_L)} - 1. \tag{4a}$$

Here $I_{0(1)}$ denotes the modified Bessel function of order 0(1). In the limit of $a/\lambda_L \ll 1$, Eq. (4a) reduces to

$$\chi_{\text{single}} \simeq -\frac{1}{8}\left(\frac{a}{\lambda_L}\right)^2. \tag{4b}$$

To simulate the realistic network of superconducting wires, COMSOL was used to determine (1) whether this quadratic variation persists for the lattice structure, and if so, (2) determine the constant of proportionality. The constant of proportionality for a network of superconducting nanowires is expected to differ from a single nanowire, owing to the enhanced effectiveness of shielding the magnetic field. In order to obtain a more complete picture of how $\chi_{network}$ is dependent on the dimensionless ratio $a/\lambda_L$, we regard $a/\lambda_L$ as a mathematical variable to be varied over two to three orders of magnitude. The simulation results, plotted in log-log scales, are shown in Fig. 5(b). It is seen that the power law behaviour of $\chi \sim (a/\lambda_L)^2$ can persist to $a/\lambda_L \sim 0.3$ before deviating from the quadratic behaviour and approaching $\chi = -1$, i.e., perfect Meissner effect, at large values of $a/\lambda_L$. The result for the same limit of $a/\lambda_L \ll 1$ is given by

$$\chi_{\text{network}} \simeq -2.5\left(\frac{a}{\lambda_L}\right)^2. \tag{4c}$$

Compared to Eq. (4b), Eq. (4c) tells us that a network of the same superconducting wires is much more effective in screening the magnetic field.

From the Ginzburg-Landau (GL) relation

$$\frac{1}{\lambda_L^2} = \frac{1}{\lambda_0^2}(1 - T/T_C), \tag{5}$$

where $\lambda_0$ denotes the magnetic penetration length at $T=0$. we obtain the expected Meissner behavior for a network of thin superconducting wires to be

$$\chi_{\text{network}} = -2.5\left(\frac{a}{\lambda_0}\right)^2 \cdot (1 - T/T_C). \tag{6}$$

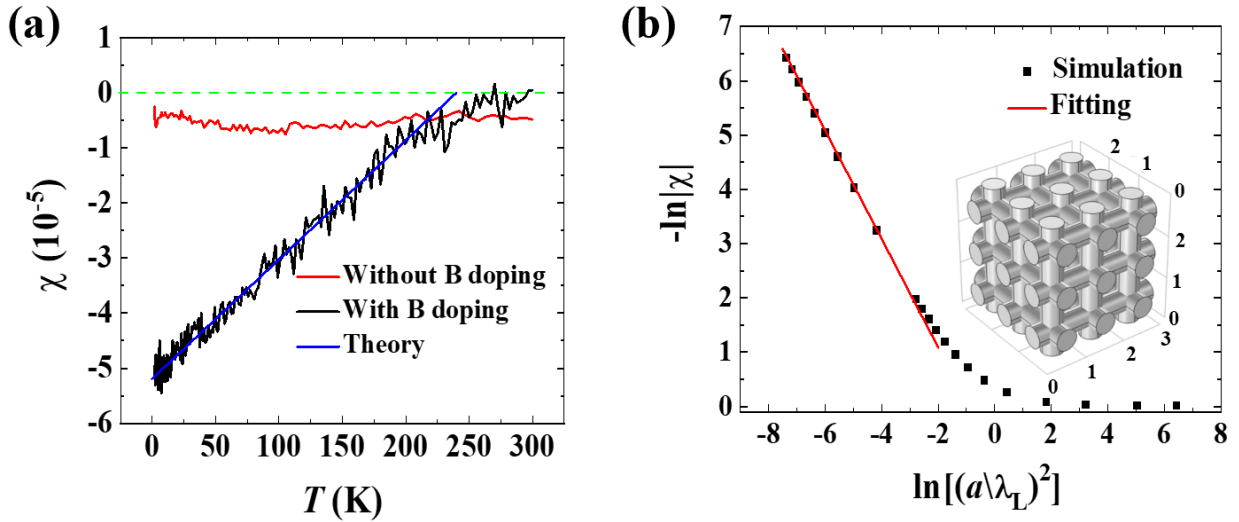

**Figure 5 (a)** Temperature dependence of magnetic susceptibility χ for with-boron-doping case (black). The blue solid line denotes the theory prediction [Eq. (6)] with $\lambda_0 = 55$ nm and $T_C = 233$ K. The data are noted to deviate from linearity at a lower temperature of ~210 K. The case without boron doping (red) is seen to be temperature independent and small in magnitude. Green dashed curve denotes χ = 0. **(b)** Simulated ln|χ| of the lattice model as a function of ln[$(a/\lambda_L)^2$]. For $(a/\lambda_L) < 0.1$, the simulated results can be described by $-\ln|\chi| = -\ln[(a/\lambda_L)^2] - \ln 2.5$. For $(a/\lambda_L) > 1$, χ is seen to approach $-1$, corresponding to the perfect Meissner effect. Inset, the illustration of the simulation model, where the length unit is nm.

In Fig. 5(a), we show the magnetic susceptibility data obtained from the SQUID measurements at a weak applied field of 30 Oe, for the two cases of with, or without, boron doping. Here the data shown were processed to remove the paramagnetic component, which always gave an upturn at low temperatures, owing to its $1/T$ behavior. Raw data and the paramagnetic signal removal procedures are given in the Appendix A. It is seen that for the case without boron doping, the magnetic susceptibility is very weakly diamagnetic and temperature-independent. However, for the case with boron doping, the linear temperature variation is exactly fitted by Eq. (6) (solid blue line), with $\lambda_0 = 220a \sim 55$ nm, where we have taken $a$=0.25 nm [35], and $T_C$=233 K. This high $T_C$ is noted to be consistent with the high transition temperature obtained from fitting the transport data with Eq. (1), as well as with the jump in the transport noise variance at around this elevated temperature, due to the appearance of the metastable states in the weak links.

In Fig. 5(b), we show the simulation results over the whole range of $a/\lambda_L$ values. The red solid line represents the behavior given by Eq. 4(c). It is seen that the complete exclusion of the magnetic field is indeed attained, i.e., the ideal Meissner effect is recovered, when $a/\lambda_L \gg 1$. The demarcation between the two regimes is seen to occur within the range of $0.3 < a/\lambda_L < 3$.

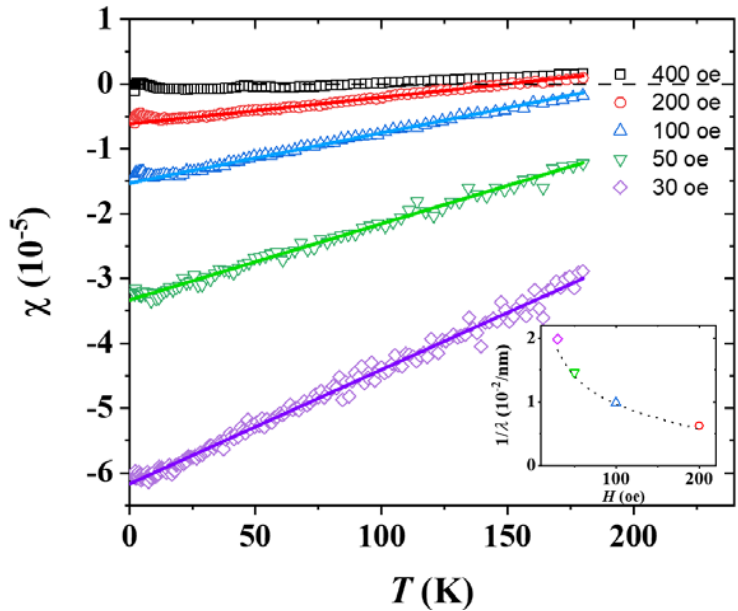

**Figure 6** Diamagnetic signals measured from a different batch of 3D networks of boron-doped CNTs.

The Meissner effect shown in Fig. 5(a) also displayed a magnetic field dependence. By using the same paramagnetic-signal-removing method described in Appendix A, the results on subtracted susceptibility measured on another sample under various magnetic fields are summarized in Fig. 6. Since that sample was actually the first sample to be measured, the measurements were programmed to stop at 180 K, as we did not expect the $T_C$ to be beyond that.

A clear linear temperature dependency of susceptibility is seen, which is consistent with the results in the main text. More importantly, with increasing magnetic field, from 30 to 400 Oe, the diamagnetic signals gradually disappear, together with a decrease of $1/\lambda$ as shown in the inset. Therefore the diamagnetic signals clearly demonstrate a magnetic field dependency, with a small critical magnetic field around 400 Oe. It should also be noted that the fitted London penetration length based on the 30 Oe data (purple) is around 50 nm, close to 55 nm for the sample shown in Fig. 5(a).

It should be noted that the Ginzburg-Landau theory is considered accurate only around the phase transition point, $T_C$. The excellent linearity of the susceptibility variation as a function of temperature, especially at low temperatures, is not an expected behaviour from the BCS theory. The deviations from the BCS theory might originate from the disorder and fluctuation effects [36,37]. However, the actual physical mechanism underlying the excellent linear temperature variation of the diamagnetic susceptibility remains a topic for future investigation.

It should also be remarked that the use of solid cylinders to model the very thin (3,0) carbon nanotube network is clearly a simplified model, intended to capture the essential difference between the Meissner effect of bulk superconductors and that of networks made of thin, hollow, atomic structures. Hence the actual values of the London penetration length obtained in our simplified model may differ from that of a more detailed, microscopic consideration.

In superconductivity, phonon spectrum is always an important consideration. For a free carbon nanotube, the inevitable low frequency floppy modes can imply that the superconducting transition temperature is not likely to be very high. In the present case, however, the geometric structure of a network comprising short carbon nanotube segments can eliminate the floppy modes, thereby making high $T_C$ superconductivity potentially possible. In addition, the zeolite template can add additional rigidity to the network.

## Concluding remarks

In summary, through both electronic transport and magnetic SQUID measurements, we have demonstrated the superconductivity in boron-doped networks of very small diameter carbon nanotubes. Owing to the untra-thin nature of the templated carbon nanotube networks, unconventional Meissner effect was experiemntally observed and theoretically explanined by applying the London equation and Ginzburg-Landau relation for the magnetic penetration length. Even though our samples are natrually inhomogeneous, the success in boron doping with diborane gas offers a relatively simple method to obtain superconducting carbon nanostructures. Our work therefore represents an opening for future research efforts to improve the sample quality and calibrate boron doping levels, so as to attain a better understanding of carbon nanotube superconductivity. In theory, it would be informative to know the solution form for the Ginzburg Landau equation in the 3D lattice network geometry. Our work also reveals the possibility of achieving high-$T_C$ superconductivity in boron-doped carbon nanostructures, which can represent a significant new development when realized.

**Author credits:** PS sustained endeavor on carbon nanotube superconductivity for the past two decades. He supervised the present research, came up with the boron doping approach, and contributed to data interpretation. JP fabricated samples and devices and carried out the experiments. BZ and YH fabricated samples and did the experiments in the initial stage of this research. TZ and XD contributed to the analytical solution and simulations of superconducting wires/networks. YW assisted in sample fabrication. NW coordinated the experiments. JP and PS wrote the manuscript.

**Acknowledgements:** PS would like to acknowledge and thank the Shun Hing Education and Charity Fund, especially the late Dr. William Mong, for sustained support of nano research. Thanks are also owed to the physics department and Jiannong Wang for allowing the use of laboratory for the duration of the experiment, after my retirement. JP would like to thank Walter Ho and Ulf Lampe for their technical assistance.

## Appendix A: Removal of paramagnetic signal

The samples for SQUID measurements are ZSM-5 zeolites with typical size in the range of 3~4 microns. To obtain the effective volume of carbon nanotube (CNT) networks, we first calculate the mass and density in the following way.

(1) <u>Mass of the CNT networks</u>. The total mass (ZSM-5 plus boron doped/non-doped CNT networks) for all the sample batches (i.e. two boron doped ones shown in Fig. 5 and Fig. 6, and non-doped one in Fig. 5) is ~ 3.5 mg (but in reality there can be some differences as we weighed only one batch). To extract the mass of the CNT networks, thermogravimetric analysis (TGA) was conducted and it was found that CNT contributes to 14.3% of the total mass. That is, in our SQUID sample the mass of CNT is around 0.5 mg.

(2) <u>Density of the CNT networks</u>. The outer radius is 0.25 nm. The density can be estimated to be $12 \cdot 1.66 \times 10^{-22} \text{g} / \left[ \pi \left( 0.25 \text{ nm}^2 \right) \cdot 0.246 \text{ nm} \right] = 0.238 \text{ g/cm}^3$.

Therefore, we can obtain the volume of CNT networks to be $2.1 \times 10^{-3} \text{cm}^3$.

For both the boron-doped and un-doped samples, the measured susceptibility always displayed an upturn at low temperatures, as shown in Fig. S1. This is very natural, due to the paramagnetic susceptibility's $1/T$ temperature dependence. We attribute its presence to the paramagnetic impurities, whose temperature dependence obeys the Curie law:

$$\chi_{para} = \beta \cdot L(\alpha / T), \qquad (\text{A-1})$$

where $L$ denotes the Langevin function $L(x) = \coth(x) - 1/x$, and the coefficients $\alpha$, $\beta$ are fitting parameters.

For the case without the boron doping, the total susceptibility is given by

$$\chi_{undoped} = \chi_{para} + C = \beta \cdot L(\alpha / T) + C . \qquad (\text{A-2})$$

The second term, a constant $C$, is intended to represent the temperature independent diamagnetic signal. Equation (A-2) can be used to fit the experimentally measured susceptibility as shown in Fig. S1(a). The fitting yields $\alpha = 3.5\text{K}, \beta = 2.1 \times 10^{-4}, C = -4.7 \times 10^{-6}$. We can see that after subtracting the paramagnetic contribution, what remains is a temperature independent diamagnetic signal.

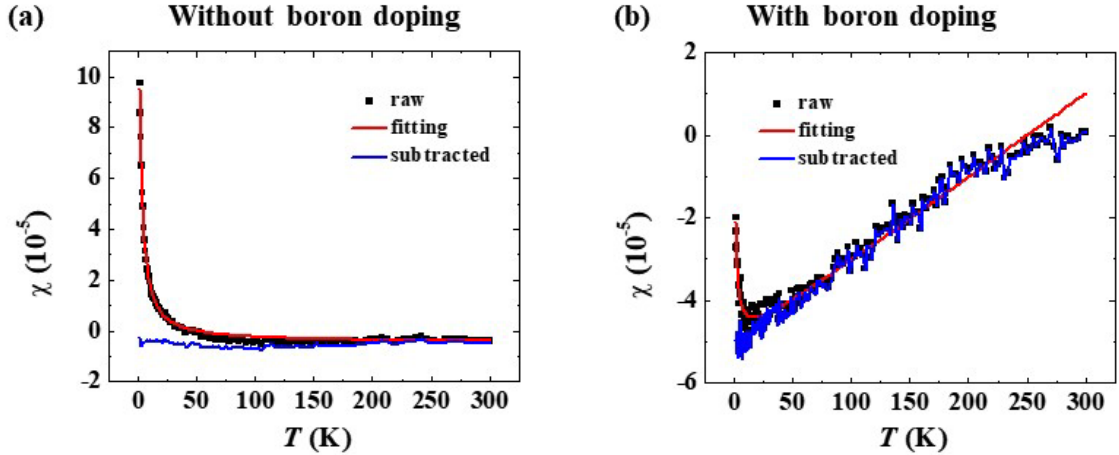

**Figure A1** Susceptibility for **(a)** without boron doping, and **(b)** with boron doping. Black squares are raw measured results in both cases. Red curves are fitted by Eq. (A-2) for **(a),** and by Eq. (A-3) for (b). Blue curves represent the results after the paramagnetic signals have been removed.

For the boron doped case, there is a linear temperature-dependent diamagnetic signal, in addition to the paramagnetic signal, that arises from the Meissner effect. Therefore, the susceptibility for the boron-doped case may be expressed as,

$$\chi_{doped} = \chi_{para} + \gamma T + C = \beta \cdot L(\alpha/T) + \gamma T + C \tag{A-3}$$

Equation (A-3) can be used to fit the experimental data shown in Fig. S1(b). The fitting yields $\alpha=1.3\text{K}, \beta=1.4\times10^{-4}, \gamma=2.23\times10^{-7}\text{K}^{-1}, C=-5.2\times10^{-5}$. We can see that after subtracting the paramagnetic signals, a linear temperature dependent diamagnetic signal is left as shown by the blue curve in Fig. S1(b). This is exactly the signal for the Meissner effect of a thin superconducting wire network in which the wire radius $a$ is much smaller than the London penetration length $\lambda_L$.

## Appendix B: Analytical solution of the London equation for a cylindrical superconductor

Here we study the solutions of the London equation

$$\nabla^2 \mathbf{H} - \frac{1}{\lambda_L^2} \mathbf{H} = 0 \tag{B-1}$$

for cylindrical superconductors. Consider the following two cases: (A) magnetic field parallel to the cylinder, and (B) magnetic field perpendicular to the cylinder.

(A) Parallel case

The London equation can be expressed in the cylindrical coordinates as

$$\frac{1}{\rho}\frac{\partial}{\partial\rho}\left(\rho\frac{\partial H}{\partial\rho}\right) + \frac{1}{\rho^2}\frac{\partial^2 H}{\partial\varphi^2} = \frac{1}{\lambda_L^2}H. \tag{B-2}$$

By applying the variable separation, i.e., $H(\rho,\varphi) = R(\rho)\Theta(\varphi)$, and considering the rotational symmetry, Eq. (B-2) can be simplified to be,

$$x^2\frac{\partial^2 R}{\partial x^2} + x\frac{\partial R}{\partial x} - x^2 R = 0, \tag{B-3}$$

where $x \equiv \rho/\lambda_L$. With the boundary condition $R(a) = H_0$, we have the solution

$$H_\parallel(\rho) = H_0 \frac{I_0(\rho/\lambda_L)}{I_0(a/\lambda_L)}. \tag{B-4}$$

where $I_n$ denotes the $n$th order modified Bessel function.

(B) Perpendicular case

When the magnetic field $H$ is perpendicular to the cylindrical axis, the external magnetic field is given by $H_0 \hat{x} = H_0 \cos\theta \hat{\rho} - H_0 \sin\theta \hat{\theta}$. The equations for the magnetic field inside the cylinder is given by

$$\begin{cases} \nabla^2 H_\rho - \dfrac{H_\rho}{\rho^2} - \dfrac{2}{\rho^2}\dfrac{\partial H_\theta}{\partial \theta} = \dfrac{H_\rho}{\lambda_L^2} \\ \nabla^2 H_\theta - \dfrac{H_\theta}{\rho^2} - \dfrac{2}{\rho^2}\dfrac{\partial H_\rho}{\partial \theta} = \dfrac{H_\theta}{\lambda_L^2} \end{cases}. \tag{B-5}$$

The rotational symmetry requires $H \sim e^{in\theta}$. By applying the transform

$$\begin{pmatrix} \tilde{H}_\rho \\ \tilde{H}_\theta \end{pmatrix} = \frac{i}{2}\begin{pmatrix} -1 & -i \\ -1 & i \end{pmatrix}\begin{pmatrix} H_\rho \\ H_\theta \end{pmatrix}, \tag{B-6}$$

we obtain a self-adjoint and diagonal differential equation

$$\nabla^2 \begin{pmatrix} \tilde{H}_\rho \\ \tilde{H}_\theta \end{pmatrix} - \frac{1}{\rho^2}\begin{pmatrix} 2n+1 & 0 \\ 0 & -2n+1 \end{pmatrix}\begin{pmatrix} \tilde{H}_\rho \\ \tilde{H}_\theta \end{pmatrix} - \frac{1}{\lambda_L^2}\begin{pmatrix} \tilde{H}_\rho \\ \tilde{H}_\theta \end{pmatrix} = 0. \tag{B-7}$$

With the similar variable separation method and boundary condition, we can solve for $\tilde{H}_{\rho(\theta)}$ to obtain the solution to Eq. (B-7) as

$$\begin{cases} H_\rho = H_0 \dfrac{I_0(\rho/\lambda_L)}{I_0(a/\lambda_L)}\cos\theta \\ H_\theta = -H_0 \dfrac{I_0(\rho/\lambda_L)}{I_0(a/\lambda_L)}\sin\theta \end{cases}. \tag{B-8}$$

From Eq. (B-8), the magnetic field inside the superconductor is seen to be parallel to the external magnetic field, as the two components have the same radial dependence.

After obtaining the solution to the London equation, the susceptibility is obtained by using the relation

$$B_{av} = (1+\chi)\mu_0 H_0 = (1+\chi)B_0, \tag{B-9}$$

where the left hand side is the volume-averaged magnetic field given by

$$B_{av} = \mu_0 H_0 \frac{\int_0^a \frac{I_0(\rho/\lambda_L)}{I_0(a/\lambda_L)} 2\pi\rho d\rho}{\pi a^2} = B_0 \frac{2I_1(a/\lambda_L)}{I_0(a/\lambda_L)} \cdot \frac{\lambda_L}{a}. \tag{B-10}$$

We obtain the susceptibility $\chi_{\text{single}}$ as

$$\chi_{\text{single}} = \frac{2I_1(a/\lambda_L)}{I_0(a/\lambda_L)} \cdot \frac{\lambda_L}{a} - 1. \tag{B-11}$$

Since the radius of the (3,0) carbon nanotubes is ultra-thin, we have $a/\lambda_L \ll 1$, and the susceptibility can be simplified to be

$$\chi_{\text{single}} = -(a/\lambda_L)^2/8. \tag{B-12}$$

## Appendix C: Critical current in the LAMH theory

For understanding the critical current based on LAMH theory, we derived the critical current as following.

The condition for the critical current is given by [9, 24]:

$$\frac{2\pi n}{(\ell/\xi)} = \frac{2\pi n}{\sqrt{1-(T/T_C)}(\ell/\xi_0)} < \frac{1}{\sqrt{3}}, \tag{C-1}$$

where $n$ is the integer winding number, $\ell$ is the sample length and $\xi$ is the coherence length. Hence the condition for the critical current is obtained by changing the inequality sign to an equality:

$$\frac{2\pi n_C}{(\ell/\xi_0)\sqrt{1-(T/T_C)}} = \frac{1}{\sqrt{3}} \tag{C-2}$$

In a superconductor, the current is given by

$$I \propto \frac{\partial \varphi}{\partial x} \sim k \sim n, \tag{C-3}$$

where $k$ denotes the wavevector. So that means

$$I_C \propto n_C = \frac{2\pi}{\sqrt{3}}(\ell/\xi_0)\sqrt{1-(T/T_C)} \tag{C-4}$$

By requiring $I_C = 0$ at ~ 40 K with $T_C$ ~ 200 K, we get an approximately linear variation of the critical current with the temperature as following,

$$I_C \sim \frac{2\pi}{\sqrt{3}}(\ell/\xi_0)\left(\sqrt{1-(T/T_C)} - \sqrt{1-40/200}\right) = I_0\left(\sqrt{1-(T/T_C)} - 2/\sqrt{5}\right) \tag{C-5}$$

The fact that the critical current vanishes at a temperature below $T_c$ is attributed to the discreteness of the integer $n_c$.

## Appendix D: Additional data on transport measurements

In this section, we provide the additional data set regarding (a) repeated *IV* curve measurements carried out at the same temperature. The purpose is to illustrate the non-repeatable nature of the fluctuations caused by the phase slips.

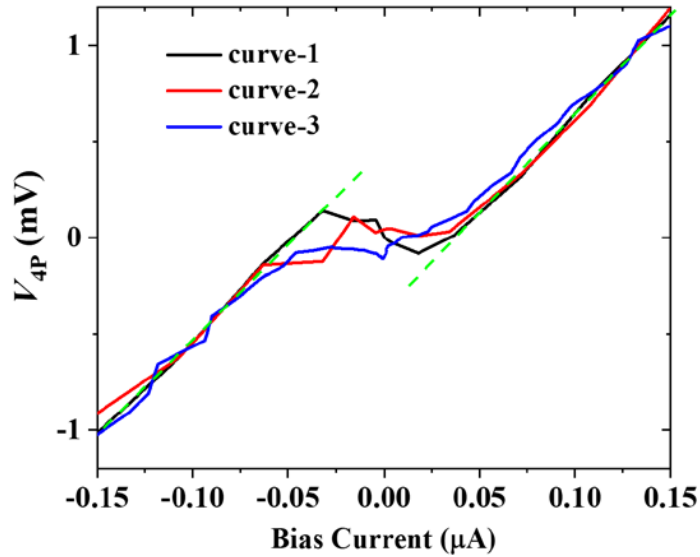

**Figure A2** Demonstrations of noise behavior by repeated *IV* curves measured at 2 K

To demonstrate that our measured *IV* curve shows large fluctuations at low temperatures, three different curve measured at 2 K are summarized in Fig. A2. Here curve 1 is used in main text, the current and voltage were averaged values over 5 measurements; while for curve 2 and 3, they were averaged over 3 times, making them a bit nosier than curve 1.